\documentclass[trackchanges,twocolumn]{aastex701}

\usepackage{amsmath}
\usepackage{mathtools}
\usepackage{gensymb}

\begin{document}

\title{Functional Analysis of Widefield Radio Interferometry}

\author[orcid=0000-0003-2064-6979,gname='Nichole',sname='Barry']{Nichole Barry}
\affiliation{School of Physics, The University of New South Wales, Australia}
\email[show]{nichole.barry@unsw.edu.au}  

\author[orcid=0000-0001-8001-0791,gname='Jaiden',sname='Cook']{Jaiden Cook}
\affiliation{CSIRO Space and Astronomy, Kensington, WA, Australia}
\email{jaiden.cook@csiro.au}

\author[orcid=0000-0001-6324-1766,gname='Cathryn',sname='Trott']{Cathryn Trott}
\affiliation{International Centre for Radio Astronomy Research, Curtin University, Bentley, WA, Australia}
\affiliation{CSIRO Space and Astronomy, Kensington, WA, Australia}
\email{cathryn.trott@curtin.edu.au}  

\author[orcid=0009-0005-6001-859X,gname='Tiasha',sname='Biswas']{Tiasha Biswas}
\affiliation{School of Physics, The University of New South Wales, Australia}
\email{tiasha.biswas@unsw.edu.au}  

\author[orcid=0009-0009-4307-5911,sname='Nazeef']{Nazeef}
\affiliation{School of Physics, The University of New South Wales, Australia}
\email{n.nazeef@unsw.edu.au}  

\begin{abstract}

Radio interferometry relies on reconstruction techniques to form images from discrete visibility measurements. As instruments encompass wider fields of view, the nuance of that reconstruction can lead to geometrical errors in the image. As a result, the estimation of the flux density of sources can change. We derive the full functional analysis of widefield radio interferometry to provide a foundation for investigating geometry-based errors. We use this framework to investigate three examples: 1) deconvolution errors resulting from a minor iteration, 2) gridding errors resulting from excluding primary beam sidelobes, and 3) $w$-term errors resulting from omitting the non-coplanar dimension on a 2\,min near-zenith transit. We use their adjoint operator descriptions to simulate each of the three cases for two fields using the Murchison Widefield Array. While the estimated errors on the flux density of reconstructed sources are small, overall errors depend on the field observed. For our two fields, we find fractional flux density errors up to 5\% due to deconvolution, 0.1\% due to gridding, and 10\% due to non-coplanarity. Our operator formalism can be used to investigate many other forms of image reconstruction errors. 

\end{abstract}

\keywords{\uat{Analytical mathematics}{38}, \uat{Astronomy image processing}{2306}, \uat{Deconvolution}{1910}, \uat{Radio interferometry}{1346}, \uat{Visibility function}{1775}, \uat{Wide-field telescopes}{1800}}

\section{Introduction}

Radio interferometers do not directly measure the sky. Instead, they measure a set of correlations, or visibilities, that represent the underlying sky brightness distribution in the frequency domain. Mapping this discrete, sampled measurement space back into the original, continuous space of the sky is a classic inverse or reconstruction problem (i.e., \citealt{bertero_linear_1985}). 

However, this inversion is becoming increasingly difficult with the advent of widefield radio interferometry because of the complexity in describing the sky sensitivity of the measurement. The Murchison Widefield Array (MWA; \citealt{tingay_murchison_2013}), the Hydrogen Epoch of Reionization Array (HERA; \citealt{deboer_hydrogen_2017}), the New Extension in Nançay Upgrading LOFAR (Nenufar; \citealt{zarka_low-frequency_2020}), the Low Frequency Array (LOFAR; \citealt{van_haarlem_lofar:_2013}), the Hydrogen Intensity and Real-time Analysis Experiment (HIRAX; \citealt{crichton_hydrogen_2022}), and the Long Wavelength Array (LWA; \citealt{ellingson_long_2009}) all have fields-of-view wider than 5 degrees on the sky. The future Square Kilometre Array Low (SKA-Low; \citealt{dewdney_square_2009}) will also boast a wide field-of-view, and will aim to do extremely sensitive image reconstruction.

This reconstruction problem is not unique to the field of radio interferometry: magnetic resonance imaging (MRI), computed tomography (CT), and seismic imaging all have incomplete or irregularly sampled measurements that are used to estimate an underlying continuous function. In order to describe their reconstruction problem, these fields have used the mathematical framework of functional analysis (i.e., \citealt{natterer_sobolev_1980,tuy_reconstruction_1981} for CT, \citealt{nemeth_least-squares_1999} for seismic imaging, and \citealt{van_de_walle_reconstruction_2000} for MRI). Functional analysis describes how functions are modified between spaces via operators, much like how rotations or projections transform ordinary vectors in geometry.

Operator-based approaches have been used in radio interferometry for a variety of physical and computational aims. For example, a wide range of work uses the operator formulism to mitigate discretization, aliasing, and non-coplanarity via numerical methods in Fourier sampling whilst maintaining efficiency \citep{wiaux_compressed_2009,mcewen_compressed_2011,dabbech_w-effect_2017,arras_efficient_2021,ye_high_2022}. A separate body of work incorporates the instrumental response in the operator framework to describe direction-dependent effects \citep{bhatnagar_correcting_2008, sullivan_fast_2012,tasse_applying_2013}, with some including the instrumental response in a covariance-weighted statistical inversion \citep{tegmark_how_1997,morales_software_2009}.

The distinction made in this work is to \textit{analytically} determine the consequences of operator approximation, rather than using operators to build new algorithms. For example, direction-dependent imaging formalisms have previously described the measurement and its adjoint to create practical reconstruction algorithms or image-plane corrections \citep{bhatnagar_correcting_2008,tasse_applying_2013}. In this work, we instead build a continuous normal operator formalism to study how it affects sky reconstruction. We separate the deterministic geometry from statistical weighting and numerical approximations to find the one natural transformation associated with the physical geometry of the sky brightness distribution, which we describe analytically through continuous functions and their transformations between spaces. 

As a contrasting example, Optimal Map-Making (OMM; \citealt{tegmark_how_1997}) solves for the optimal reconstruction operator given a noise covariance. However, this extends beyond the underlying natural transformation based on geometric principles alone, which does not depend on the noise. Therefore, a statistically optimal reconstruction given by OMM will not necessarily result in a geometrically faithful reconstruction of the sky. Practical choices such as covariance weighting, discretizations, compact support, or simplified beam models may be scientifically or computationally useful, but they modify the natural geometry of the problem, which redefines distances, correlations, and orthogonality.

In this paper, we use the functional analysis framework to describe the errors associated with this departure from the true geometry, specifically in image reconstruction using widefield radio interferometers. In particular, we describe errors arising from misrepresenting the physical instrumental response, independent of errors associated with numerical Fourier interpolation or statistical weighting. This provides an analytic framework for identifying and estimating reconstruction errors separately from the particular imaging algorithm used to realize the operator.

In Section~\ref{sec:now_math}, we will derive gridding from first principles using functional analysis, specifically for the coplanar case for simplicity. We then explore the impact of the normal operator in Section~\ref{sec:operator_action} via the example of using minor cycles, or non-adjoint driven reconstruction, inside cycles of adjoint-driven major cycles. Using the linearity of operators previously introduced, Section~\ref{sec:gridding_kernel} then describes the errors associated with the choice of gridding kernel. We reintroduce the w-term in Section~\ref{sec:math_with_w} and rederive gridding using functional analysis to show how this linear formalism can easily be modified. In Section~\ref{sec:w_impact}, we continue our exploration of the impact of errors in this operator formalism, specifically with regard to the w-term. We outline a number of possibilities for future work using our functional analysis derivation in Section~\ref{sec:SH_basis_future}, and conclude in Section~\ref{sec:conclusions}.

\section{Functional Analysis derivation of gridding \textit{without w}} 
\label{sec:now_math}

We will describe the function that represents the act of taking visibility measurements as the \textit{operator}. This operator is responsible for transforming our geometry from a continuous, square-integrable function space, otherwise known as a \textit{Hilbert space} $L^2(\mathbb{R}^2)$, to a sampled measurement space represented through delta distributions, otherwise known as a \textit{distribution space} $D'(\mathbb{R}^2)$. 

The function that represents the act of using those visibility measurements to make an estimate of the sky brightness distribution is called the \textit{adjoint operator}, or simply the adjoint. The combined resulting distortion to the sky brightness distribution caused by applying the operator and adjoint operator is called the \textit{normal operator}, and it encodes the result of the combined operator action. 

We begin our derivation with the visibility function, $V$, defined as the Fourier response of an interferometer to the sky brightness. The continuous radio interferometric measurement equation for Stokes I is

\begin{multline}
\label{eq:RIME}
    V(u,v,w) = \iint A(l,m) I(l,m) \\ e^{-i2\pi(ul +vm + w(\sqrt{1-l^2-m^2}-1))} \frac{\textrm{d}l\textrm{d}m}{\sqrt{1-l^2-m^2}},
\end{multline}
where $(l,m,n)$ are the direction cosines on the sky and $(u,v,w)$ are their Fourier-dual coordinates \citep{hamaker_understanding_1996}. The instrumental sensitivity response as a function of sky-location, otherwise known as the primary beam $A(l,m)$, is included as one simple term, but can be expanded in Jones or Mueller matrices to take into account polarisation properties, as well as other sky polarisations \citep{smirnov_revisiting_2011, tasse_applying_2013}. $I(l,m)$ is the Stokes I flux density on the sky, and $1/\sqrt{1-l^2-m^2}$ is the Jacobian determinant. In this equation, $n$ has explicitly been represented as a function of $(l,m)$, which is possible because $(l,m,n)$ lie on the hemisphere of the curved 2D celestial sky and are thus not independent.

\citet{thompson_interferometry_2017} defines $A(l,m)$ as the normalized reception pattern on the sky. In this simplified, unpolarized formalism, $A(l,m)$ is the complex product of the normalized voltages from each element. If there are differences between the voltage phases, this will combine with the rest of the phases from the signal chain (i.e., via amplifiers and transmission lines) and will be accounted for with a calibration reference phase. Therefore, we will leave $A(l,m)$ as complex, even though it is a common algorithmic choice to make it real-valued. 

Now, Equation~\ref{eq:RIME} for a real, discrete interferometer is actually a sampled function at the $(u,v,w)$ points with measurements. This discreteness can be represented as the operator $S(u,v,w)$, which is a sum of delta functions at discrete, measured points \citep{cornwell_radio-interferometric_1992,bracewell_fourier_1999}. To simplify Equation~\ref{eq:RIME}, we can also extract the $w$ component of the exponential into its own term: $W(l,m;w)$. 

This gives an observed visibility distribution function,
\begin{multline}
\label{eq:sampled_v_w}
V_{\mathrm{obs}}(u,v,w) = S(u,v,w) \iint A(l,m) I'(l,m) \\ W(l,m;w) e^{-i2\pi(ul +vm)} \textrm{d}l\textrm{d}m,
\end{multline}
where we have absorbed the Jacobian determinant into the sky brightness $I'(l,m)$.

For the case where the interferometer is fully coplanar and observing zenith, $w$ is 0 and the resulting $W(l,m;w)$ term is 1. Incorporating this simplification, along with representing $S(u,v,w)$ as the sum of delta functions, results in
\begin{multline}
\label{eq:sampling}
V_{\mathrm{obs}}(u,v,w=0) = \\ \sum_{n=1}^N \left[\iint A(l,m) I'(l,m) e^{-i2\pi(ul +vm)} \textrm{d}l\textrm{d}m\right] \\ \delta(u-u_n)\delta(v-v_n),
\end{multline}
and the convolution theorem gives
\begin{multline}
\label{eq:conv_at_point}
V_{\mathrm{obs}}(u,v,w=0) = \\ \sum_{n=1}^N \left[(\tilde{I'} * \tilde{A})(u,v)\right] \delta(u-u_n)\delta(v-v_n),
\end{multline}
where the Fourier transform of $I'$ and $A$ is denoted by $\tilde{I'}$ and $\tilde{A}$.

It is important to note that this is a distribution of a finite set of point evaluations of a function, or a distribution-valued functional. We can describe Equation~\ref{eq:conv_at_point} as a distribution of a finite set of points\footnote{Using the sifting property of delta functions, i.e., \\ $f(x,y)\delta(x-x_n)\delta(y-y_n) = f(x_n,y_n)\delta(x-x_n)\delta(y-y_n)$},
\begin{multline}
\label{eq:scalar_conv_at_point}
V_{\mathrm{obs}}(u,v,w=0) = \\ \sum_{n=1}^N (\tilde{I'} * \tilde{A})(u_n,v_n) \delta(u-u_n)\delta(v-v_n).
\end{multline}

Via the act of taking measurements, we have taken a function living in Hilbert space (the visibility function, Equation~\ref{eq:RIME}) and mapped it to distribution space (discrete observed visibilities, Equation~\ref{eq:scalar_conv_at_point}). We can attempt to reverse this operation through a process called gridding. For an operator $O : L^2(\mathbb{R}^2) \rightarrow D'(\mathbb{R}^2)$, we aim to find an adjoint operator $O^\dagger : D'(\mathbb{R}^2) \rightarrow L^2(\mathbb{R}^2)$.  

This desire to map back to Hilbert space using $O^\dagger : D'(\mathbb{R}^2) \rightarrow L^2(\mathbb{R}^2)$ actually restricts the original operator. If the original operator was just the sampling of the continuous $V(u,v,w)$, then the adjoint would be an operator that attempts to undo the act of sampling. Once a collapse of an infinite space to a finite space via delta functions occurs, it cannot be undone. The reverse mapping is a series of delta functions on a continuous space, which is the definition of distribution space, $O^\dagger : D'(\mathbb{R}^2) \rightarrow D'(\mathbb{R}^2)$. Hilbert space cannot be achieved. 

If, however, we act the original operator on $\tilde{I'}(u,v,w)$ such that the action is a \textit{convolution and sampling}, we can map back to Hilbert space. Attempting to reverse the convolution spreads the delta functions back into a continuous Hilbert space. While the act of sampling is never truly invertible, the act of convolution and sampling is adjointable, allowing for $O^\dagger : D'(\mathbb{R}^2) \rightarrow L^2(\mathbb{R}^2)$. Thus, the adjointability of our measurements is solely due to the presence of the primary beam response, $A(l,m)$.

The adjoint operator must maintain linearity, preserve location, and reproduce results for all functions. This is summarized by the identity $\langle O(F),\phi\rangle=\langle F,O^\dagger(\phi) \rangle$ for any generalized function $F$ and any test function $\phi$. Our continuous function is $\tilde{I'}(u,v)$ as discussed prior, and a suitable and simple test function is a 2D delta function, giving the LHS of the identity as $\langle O(F),\phi\rangle= \langle O(\tilde{I'}(u,v)), \delta_{u_m}\delta_{v_m} \rangle$. Our choice is very deliberate; the forward operator acting on the sky brightness, $O(\tilde{I'})$ gives us our discrete observed visibilities (Equation~\ref{eq:scalar_conv_at_point}) exactly, which is collapsed via the delta functions in our choice of test function. Thus, the LHS of the adjoint operator identity gives
\begin{multline}
\label{eq:LHS}
\langle\sum^N_{n=1} (\tilde{I'} * \tilde{A})(u_n,v_n) \delta_{u_n}\delta_{v_n},\delta_{u_m}\delta_{v_m}\rangle = 
(\tilde{I'} * \tilde{A})(u_m,v_m) \\ = \iint \tilde{I'}(u,v)\tilde{A}(u_m-u,v_m-v)\textrm{d}u\textrm{d}v,
\end{multline}
where we have used the notation $\delta_{u_n}=\delta(u-u_n)$ and the orthogonality relation of the Kronecker delta $\langle\delta_n,\delta_m\rangle = \delta_{nm}$. Given our deliberate choices of $F$ and $\phi$, this now must equal the RHS of our adjoint operator identity,
\begin{multline}
\label{eq:RHS}
\langle \tilde{I'}(u,v) ,O^\dagger(\delta_{u_m}\delta_{v_m})\rangle = \\ \iint \tilde{I'}(u,v) [O^\dagger(\delta_{u_m}\delta_{v_m})]^* \textrm{d}u\textrm{d}v.
\end{multline}

Equating Equations~\ref{eq:LHS} and \ref{eq:RHS}, the adjoint operator identity is only valid if and only if 
$[O^\dagger(\delta_{u_m}\delta_{v_m})]^* = \tilde{A}(u_m-u,v_m-v)$. Multiplying both sides by a conjugate and using the kernel identity $A^\dagger(x - x_n) = A^*(x_n-x)$, we get that the adjoint operator is 
\begin{equation}
\label{eq:adjoint}
O^\dagger(\delta_{u_m}\delta_{v_m}) = \tilde{A}^\dagger(u-u_m,v-v_m).
\end{equation}
This is our gridding operator. At a given point, this operator is the shifted Hermitian conjugate of the $(u,v)$ space normalized reception pattern, or primary beam. Again, $O^\dagger$ maps our distribution space back to Hilbert space; thus, the subtlety of the adjoint operator being defined point-by-point as $O^\dagger(\delta_{u_m}\delta_{v_m})$ is due to its distribution-space origins. 

To perform gridding, we apply the adjoint operator to the observed visibilities to obtain a mapping of the Fourier-space sky, $(O^\dagger O)(\tilde{I'})$. Given that the original operator action, $O(\tilde{I'})$, resulted in our observed visibility distribution $V_{\mathrm{obs}}$ in Equation~\ref{eq:scalar_conv_at_point}, gridding is simply $O^\dagger(V_{\mathrm{obs}})$. We can take advantage of the linearity of operators to act the adjoint operator on any part of Equation~\ref{eq:scalar_conv_at_point}. Thus, we have 
\begin{multline}
\label{eq:gridding}
O^\dagger(V_{\mathrm{obs}}) = \\ \sum_{n=1}^N (\tilde{I'} * \tilde{A})(u_n,v_n) O^\dagger(\delta(u-u_n)\delta(v-v_n)) = \\ \sum_{n=1}^N V_{\mathrm{obs}}(u_n,v_n) \tilde{A}^\dagger(u-u_n,v-v_n).
\end{multline}
We have deliberately chosen to act the adjoint operator on the delta functions, allowing us to use our definition of the gridding operator in Equation~\ref{eq:adjoint}.
Equation~\ref{eq:gridding} is our attempt to map our measurements to continuous Hilbert space; however, this does not mean that Equation~\ref{eq:gridding} is an estimate of the continuous function just yet. 

In order to estimate the continuous $\tilde{I'}(u,v,w)$, we must compensate for the total action of the operators, known as the normal operator, $O^\dagger O:L^2(\mathbb{R}^2) \rightarrow L^2(\mathbb{R}^2)$. The first operator, $O$, convolved and sampled the true continuous distribution, and the second operator, $O^\dagger$, gridded the samples. Importantly, $O^\dagger O \neq I$, thus the operation of measuring data and then mapping back to a Hilbert space results in something akin to a distortion. We can undo this distortion by applying the inverse, $(O^\dagger O)^{-1}$, to our gridded visibilities,
\begin{multline}
\label{eq:reverse_operation}
(O^\dagger O) (\tilde{I'}) = O^\dagger(V_{\mathrm{obs}}) \\ \xrightarrow{(O^\dagger O)^{-1}} \tilde{I'} = (O^\dagger O)^{-1} O^\dagger(V_{\mathrm{obs}})
\end{multline}

To calculate the distortion, we apply the operators to a generalized function $F$ in Hilbert space. Application of the first operator, $O$, gives the generalized result of our observed visibility distribution (Equation~\ref{eq:scalar_conv_at_point}). Application of the second operator, $O^\dagger$, gives the generalized result of our gridded visibilities (Equation~\ref{eq:gridding}). We can highlight the combined effect of the operations by expanding the convolution via integration, and moving around the sum,
\begin{multline}
\label{eq:generalised_combined}
(O^\dagger O) (F) = \sum_{n=1}^N (F * \tilde{A})(u_n,v_n) \tilde{A}^\dagger(u-u_n,v-v_n) = \\ \sum_{n=1}^N \iint F(u^{\prime\!},v^{\prime\!}) \tilde{A}(u_n-u',v_n-v') \textrm{d}u^{\prime\!} \textrm{d}v^{\prime\!} \tilde{A}^\dagger(u-u_n,v-v_n) \\ = \iint F(u^{\prime\!},v^{\prime\!}) K(u,v;u',v') \textrm{d}u^{\prime\!} \textrm{d}v^{\prime\!},
\end{multline}
with the kernel defined as
\begin{multline}
\label{eq:generalised_kernel}
K(u,v;u',v') = \\ \sum_{n=1}^N \tilde{A}(u_n-u',v_n-v')  \tilde{A}^\dagger(u-u_n,v-v_n).
\end{multline}
This is the definition of an integral operator with a four-dimensional kernel. Practically speaking, it is the sum of the shifted outer product of the primary beam response for each measurement in both $(u,v)$ and $(u',v')$ space; it lives in 4D. Unfortunately, this is where the purely analytical version of our gridding functional analysis ends -- this integral operator is not invertible, and thus we cannot calculate the distortion correction, $(O^\dagger O)^{-1}$. The kernel is neither separable nor shift-invariant, and is thus not generally invertible in closed form. From a matrix point of view, the operator normal $(O^\dagger O)^{-1}$ is not fully independent when there is missing $(u,v)$ coverage in the measurement -- it is not full-rank and therefore not invertible \citep{dillon_fast_2013,trott_chips:_2016}. We will explore the impact of these ideas further in Section~\ref{sec:operator_action}. 

Direct acknowledgment of the distortion of the operator is rarely written explicitly in the literature, though it can be implicitly handled in generalized correction operations, including some deconvolution approaches that can account for the measurement operator. The idea of an inverse operator correction appears in \citet{bhatnagar_correcting_2008}, \citet{morales_software_2009}, \citet{tasse_applying_2013}, \citet{dillon_fast_2013}, \citet{sullivan_fast_2012}, and \citet{trott_chips:_2016}, and can sometimes be referred to as a deconvolution operator.

We have used an adjoint-operator derivation to motivate gridding in a functional analysis framework, explicitly starting from a continuous domain, moving into a sampled domain, and returning back again to a continuous domain. We explicitly construct the measurement operator and its adjoint in a continuous setting, highlighting the role of the instrumental sensitivity response as the adjoint kernel and making explicit the resulting non-trivial distortion caused by the operator $O^\dagger O$. This derivation is most similar in concept to \citet{bhatnagar_correcting_2008} and \citet{ye_high_2022}, but with explicitly including the instrumental sensitivity response -- an important aspect given that it becomes the adjoint operator, and the sole reason why there is adjointability of the inverse reconstruction at all. 

\section{The Impact of the normal operator}
\label{sec:operator_action}

A reconstruction of the continuous sky brightness distribution from a discrete set of visibilities will naturally be incomplete. In radio astronomy terms, this incompleteness gives rise to the image-based point-spread function (PSF), which is generally mitigated via a deconvolution algorithm. This contamination is a result of the normal operator in our functional analysis derivation, and it can have a non-trivial impact on the resulting reconstruction.

The normal operator is the combined action of the measurement operator and its adjoint, or $O^\dagger O$, and its resulting impact is very different for wide versus narrow instrumental sensitivity response.
To illustrate this point, we calculate the Fourier transform of the uncorrected sky with distortion, $I_D(l,m) \coloneqq\mathcal{F}^{-1}_{u,v}\!\left[ O^\dagger(V_{\mathrm{obs}})(u,v) \right]$, otherwise known as the ``dirty image." Using the normal operator on the sky brightness (via Equation~\ref{eq:generalised_combined}) to expand $O^\dagger(V_{\mathrm{obs}})$,
\begin{multline}
I_D(l,m) = \mathcal{F}^{-1}_{u,v} \left[ \iint \tilde{I'}(u',v') K(u,v;u',v') \textrm{d}u' \textrm{d}v' \right] \\ = \iint \tilde{I'}(u',v') \mathcal{F}^{-1}_{u,v}[K(u,v;u',v')] \textrm{d}u' \textrm{d}v',
\end{multline}
where the Fourier transform only acts on the kernel because it is the only part of the equation which explicitly depends on $(u,v)$. To get to the desired sky coordinate space, we expand $\tilde{I'}(u,v)$ as its Fourier transform, giving 
\begin{multline}
\label{eq:dirty_sky}
I_D(l,m) = \iint_{l^{\prime\!},m^{\prime\!}} I'(l^{\prime\!},m^{\prime\!}) \\ \left( \iint_{u^{\prime\!},v^{\prime\!}} e^{-i2\pi(u^{\prime\!}l^{\prime\!}+v^{\prime\!}m^{\prime\!})} \mathcal{F}^{-1}_{u,v}[K(u,v;u^{\prime\!},v^{\prime\!})] \textrm{d}u^{\prime\!} \textrm{d}v^{\prime\!} \right)\textrm{d}l^{\prime\!} \textrm{d}m^{\prime\!} \\ = \iint I'(l',m') \mathcal{F}_{u^{\prime\!},v^{\prime\!}}\mathcal{F}^{-1}_{u,v}[K(u,v;u',v')] \textrm{d}l' \textrm{d}m'.
\end{multline}

The mapping between the true sky $I'$ and the dirty sky $I_D$ is an integral operator, $\mathcal{F}_{u^{\prime\!},v^{\prime\!}}\mathcal{F}^{-1}_{u,v}[K(u,v;u',v')]$. Essentially, this kernel describes how each point on the sky is coupled to every other point on the sky, which depends on the combined sampling-instrument response. Thus, the PSF is position-dependent; it is not a simple convolution operator. The discrete analog of this integral operator is the holographic mapping function in \citet{sullivan_fast_2012}. 

Classical deconvolution in algorithms like CLEAN \citep{hogbom_aperture_1974,schwab_relaxing_1984,cornwell_multiscale_2008} undo a convolution by a global PSF, and were designed for narrow-beam instrumentation. Instruments with a nearly flat primary beam, $A(l,m)$, will correspondingly have a narrow, delta-like $\tilde{A}(u,v)$. In the limit of $\tilde{A}(u,v) \rightarrow \delta$, the normal operator on a generalized function (Equation~\ref{eq:generalised_combined}) becomes
\begin{multline}
\label{eq:generalised_combined_limit}
(O^\dagger O) (F) \xrightarrow{\tilde{A}=\delta}  \sum_{n=1}^N F(u_n,v_n) \delta(u-u_n)\delta(v-v_n) =  \\ \iint F(u',v') K(u,v; u',v')\big|_{\tilde{A}=\delta} \textrm{d}u' \textrm{d}v',
\end{multline}
with the kernel defined as
\begin{multline}
\label{eq:generalised_kernel_limit}
K(u,v;u',v')\big|_{\tilde{A}=\delta} = \\ \sum_{n=1}^N \delta(u'-u_n)\delta(v'-v_n)\delta(u-u_n)\delta(v-v_n).
\end{multline}
The action of the combined operators in this limit is just a sampling. The kernel is diagonal, and thus is now effectively 2D and invertible. In this limit, the integral operator on the true sky to form the dirty sky (Equation~\ref{eq:dirty_sky}) is
\begin{multline}
\label{eq:global_convolution_operator}
\mathcal{F}_{u^{\prime\!},v^{\prime\!}}\mathcal{F}^{-1}_{u,v}K(u,v;u',v') = \\  \sum_{n=1}^N\mathcal{F}_{u^{\prime\!},v^{\prime\!}}\left[\delta(u^\prime\!-u_n)\delta(v^\prime\!-v_n)\right] \mathcal{F}^{-1}_{u,v}\left[\delta(u-u_n)\delta(v-v_n)\right] \\ = \sum_{n=1}^N e^{-2\pi i (u_nl'+v_nm')} e^{2\pi i (u_nl+v_nm)}.
\end{multline}
This integral operator has the form $K(l-l',m-m')$. The dirty sky in Equation~\ref{eq:dirty_sky} can therefore be represented as a convolution, $I'(l,m) * \sum_{n=1}^N e^{2\pi i (u_nl+v_nm)}$, where the convolution is with a point spread function or synthesized beam dependent on the array layout. The narrow-beam limit has a global image distortion that is purely based on the measurement $(u,v)$ coverage. 

In the limit $\tilde{A}(u,v) \rightarrow \delta$, image reconstruction is an analytically solvable problem. The normal operator is invertible and global. However, it is difficult to achieve this limit in practice. Using the linearity of operators, we can quantify the level of error present when a global PSF is assumed for a practical, widefield instrument. 

We can split our instrumental sensitivity into two parts: one which will result in a global average PSF, and one which will result in differences from the global average PSF,
\begin{equation}
A(l,m) = A_G(l,m) + A_{\Delta G}(l,m).
\end{equation}
The convolution and point evaluation of our observed visibility distribution (Equation~\ref{eq:scalar_conv_at_point}) is linear, and thus we can write 
\begin{multline}
V_{\mathrm{obs}}(u,v,w=0) = \\ \sum_{n=1}^N \big[(\tilde{I'} * \tilde{A}_G)(u_n,v_n)  +(\tilde{I'} * \tilde{A}_{\Delta G})(u_n,v_n)\big] \\ \delta(u-u_n)\delta(v-v_n).
\end{multline}
Using the same logic as Section~\ref{sec:now_math}, we can solve for the adjoint operator,
\begin{multline}
\label{eq:global_operator_adjoint}
O^\dagger = \tilde{A}^\dagger_G(u-u_n,v-v_n) + \tilde{A}^\dagger_{\Delta G}(u-u_n,v-v_n) \\ = O^\dagger_G + O^\dagger_{\Delta G}.
\end{multline}
The operator and its adjoint are simply a linear combination of the global average operator and the local difference operator. The full normal operator, using this linear combination, is now
\begin{multline}
\label{eq:global_operators}
O^\dagger O = O^\dagger_G O_G + O^\dagger_G O_{\Delta G} + O^\dagger_{\Delta G} O_G + O^\dagger_{\Delta G} O_{\Delta G}.
\end{multline}
To assess the consequences of a simple deconvolution, we will have to make an assumption about the inverse action that it applies. In the general sense, deconvolution algorithms iteratively solve for the true sky via a form of least-squares inversion, which can be described as a pseudoinverse. In the optimistic sense, the aim of a simple deconvolution is to remove $(O^\dagger_G O_G)^{-1}$ in the image space by reversing a convolution of the integral operator of the form $K(l-l',m-m')$ (Equation~\ref{eq:global_convolution_operator}).

Using the definition $I_D(l,m) \coloneqq\mathcal{F}^{-1}_{u,v}\!\left[ O^\dagger(V_{\mathrm{obs}})(u,v) \right] = \mathcal{F}^{-1}_{u,v}\!\left[ O^\dagger O(\tilde{I'})\right]$, we can build an approximation for the effects on the dirty image. The dirty image, decomposed into global and non-global operator contributions, is
\begin{multline}
I_D = \mathcal{F}^{-1}_{u,v}\left[O_G^\dagger O_G(\tilde{I'}) \right] + \\ \mathcal{F}^{-1}_{u,v}\left[ (O_G^\dagger O_{\Delta G} + O_{\Delta G}^\dagger O_{G} + O_{\Delta G}^\dagger O_{\Delta G})(\tilde{I'})  \right].
\end{multline}
In order to deconvolve in image space, we multiply in Fourier space. After applying $\mathcal{F}_{u,v}$, and multiplying by $(O^\dagger_G O_G)^{-1}$, we have
\begin{multline}
\label{eq:image_diff_decon_intermediate}
\tilde{I}_D = (O_G^\dagger O_G)^{-1}(O_G^\dagger O_G)(\tilde{I'}) + (O_G^\dagger O_G)^{-1} \\ (O_G^\dagger O_{\Delta G} + O_{\Delta G}^\dagger O_{G} + O_{\Delta G}^\dagger O_{\Delta G})(\tilde{I'}).
\end{multline}
Again, this simple deconvolution step aims to isolate $\tilde{I'}$, and for the first term, it has successfully done so. However, the second term in Equation~\ref{eq:image_diff_decon_intermediate} remains. This difference in the expectation from reality in the image space is
\begin{multline}
\label{eq:image_diff_decon}
I_D - I' = \mathcal{F}^{-1}_{u,v}\left[(O_G^\dagger O_G)^{-1} \right.\\ \left. (O_G^\dagger O_{\Delta G} + O_{\Delta G}^\dagger O_{G} + O_{\Delta G}^\dagger O_{\Delta G})(\tilde{I'})\right].
\end{multline}
This is the residual in a deconvolved image using a simple, global deconvolution algorithm -- it arises from applying the incorrect normal operator inverse to an image which has been distorted by a non-flat instrumental response. 

\subsection{Simulated example of deconvolution errors}
\label{subsec:decon_example}

To give an example of this non-global operator loss and its effects on the image, we simulate two zenith-pointed MWA Epoch of Reionisation observations using the GLEAM catalog \citep{hurley-walker_GLEAM} and the WODEN simulator \citep{line_woden_2022}. One field does not have bright sources\footnote{Phased to Right Ascension (RA) $= 0.00^\mathrm{h}$ and Declination (Dec) $= -27\degree$, with the Sculptor Galaxy (NGC 253) being the brightest source in the main field-of-view at approximately 17\,Jy at 200\,MHz.} and the other field has a very bright source right outside the half-power point of the primary beam\footnote{Phased to Right Ascension (RA) $= 60^\mathrm{h}$ and Declination (Dec) $= -30\degree$, with Fornax A being the brightest source in the field at approximately 750\,Jy at 154\,MHz \citep{mckinley_modelling_2015}.}. However, as shown in \citet{barry_aliasing_2024}, our own Galaxy is setting over the horizon for the MWA at approximately 10,000\,Jy at 200\,MHz, causing an observation-wide Gibbs-sampling ringing. We simulate this component with a catalog from the Engineering Development Array 2 \citep{kriele_imaging_2022} and the WODEN simulator. 

Our simulation spans 30.72\,MHz centered on approximately 180\,MHz, with 80\,kHz spectral resolution, and runs over 2\,min, with 2\,sec timing resolution. Our simulation is a Phase I layout (pseudo-random), pointed straight up to phase center, with a flat altitude across the array to ensure a coplanar response in accordance with the $w=0$ assumption thus far. There are no additional contributions to the simulation: it consists only of the sky $I$ and the instrumental response $A$.

We assess the effects of non-global operator contributions on the reconstructed image using Fast Holographic Deconvolution (FHD; \citealt{sullivan_fast_2012}). FHD constructs the discretized version of the normal operator in the ``holographic mapping function" and applies it in every deconvolution cycle\footnote{A cycle determines how many flux density components are fit to the image before they are removed.}. This is in contrast to most other deconvolution algorithms, which consist of minor (simple) and major (complex) cycles. Minor cycles generally apply the global convolution operator only, and major cycles apply the full operator after a set of minor cycles in an effort to balance accuracy and efficiency. We can diagonalize the holographic mapping function in FHD to emulate a minor cycle ($O^\dagger O = O^\dagger_G O_G$), and use the full holographic mapping function in FHD to emulate a major cycle ($O^\dagger O = O^\dagger_G O_G + O^\dagger_G O_{\Delta G} + O^\dagger_{\Delta G} O_G + O^\dagger_{\Delta G} O_{\Delta G}$).

For our control, we run FHD using just major cycles, or the full normal operator deconvolution, to capture 30,000 components across the entire sky. This captures about 50\% of the total flux of the sky, with the rest of the flux estimated with a simple CLEAN bias correction. We then compare to a run where 10s of minor cycles, or global-only normal operator deconvolution, are performed between each major cycle for the same number of total components. Deconvolution components are then grouped into sources and cross-matched with the input catalog. Sources are ``restored" with a CLEAN bias correction and a Gaussian synthesized beam to a residual image. 

\begin{figure}
\centering
	\includegraphics[width =\columnwidth]{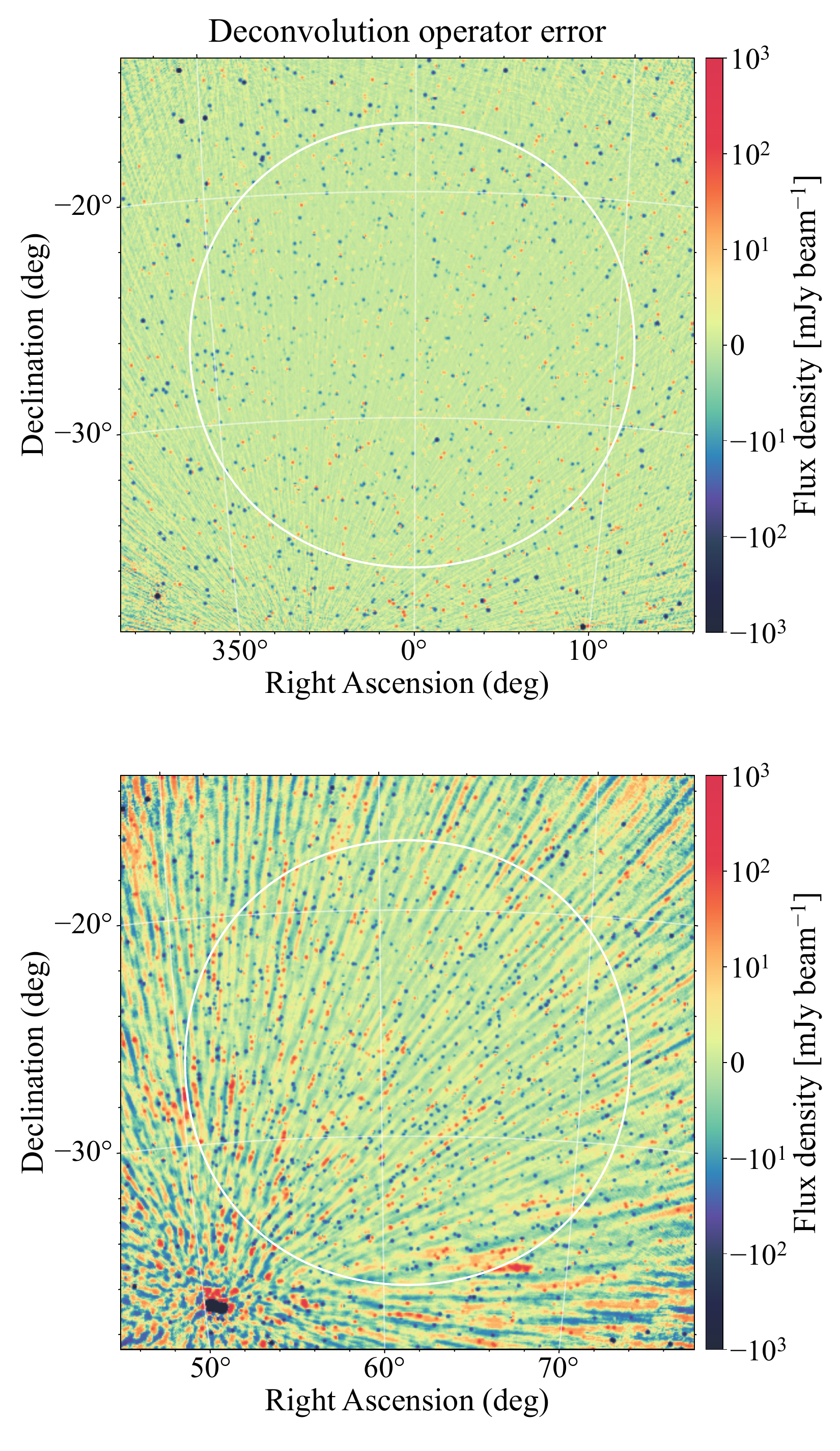}
	\caption{The difference in restored images for a deconvolution with 10's of minor cycles (global-only normal operator applied) for every major cycle (full normal operator applied) compared to restored images for a deconvolution that always performs major cycles. The top panel is a simulation with no bright sources, and the bottom panel is a simulation with Fornax A at RA$\approx$50$\degree$. Red indicates that there is more flux density in the deconvolution that uses global-only normal operators for minor cycles, and blue indicates there is less. Recovered flux densities are biased by their background residuals. Residuals have been scaled down 97\% for visual purposes, and the half-power beam level is shown in white.}
	\label{fig:decon_flux_error}
\end{figure}

Figure~\ref{fig:decon_flux_error} is the difference between the restored image made with major and minor cycles, or combinations of $O^\dagger O = O^\dagger_G O_G + O^\dagger_G O_{\Delta G} + O^\dagger_{\Delta G} O_G + O^\dagger_{\Delta G} O_{\Delta G}$ and $O^\dagger O = O^\dagger_G O_G$ for efficiency purposes, and the restored image made with only major cycles, or $O^\dagger O = O^\dagger_G O_G + O^\dagger_G O_{\Delta G} + O^\dagger_{\Delta G} O_G + O^\dagger_{\Delta G} O_{\Delta G}$. Residuals have been scaled down 97\% for visual purposes. Blue indicates that the major/minor cycle deconvolution has less flux in the restored image, and red indicates that it has more flux. The half-power beam level is highlighted in white. 

In the top panel, the input simulation does not have any significantly bright sources. The resulting difference in recovered brightness of sources between the two runs is consistent with 0 and shows no significant bias when compared to the input catalog. 

In contrast, the bottom panel has significant differences that are dependent on the background residuals. Fornax A, the bright source with complex morphology at RA$\approx$50$\degree$, is not captured well during minor cycles. In fact, flux density is captured so poorly when minor cycles are a part of the deconvolution that the residuals are greater than the catalog flux density, resulting in large amounts of PSF structure that streak across the entire image. There is approximately 20\% more false sources and 15\% fewer cross-matched sources in the major/minor cycle deconvolution than in the major deconvolution. By eye, the positive residual differences correspond to positive source flux density differences, and vice versa. 

\begin{figure}
\centering
	\includegraphics[width =\columnwidth]{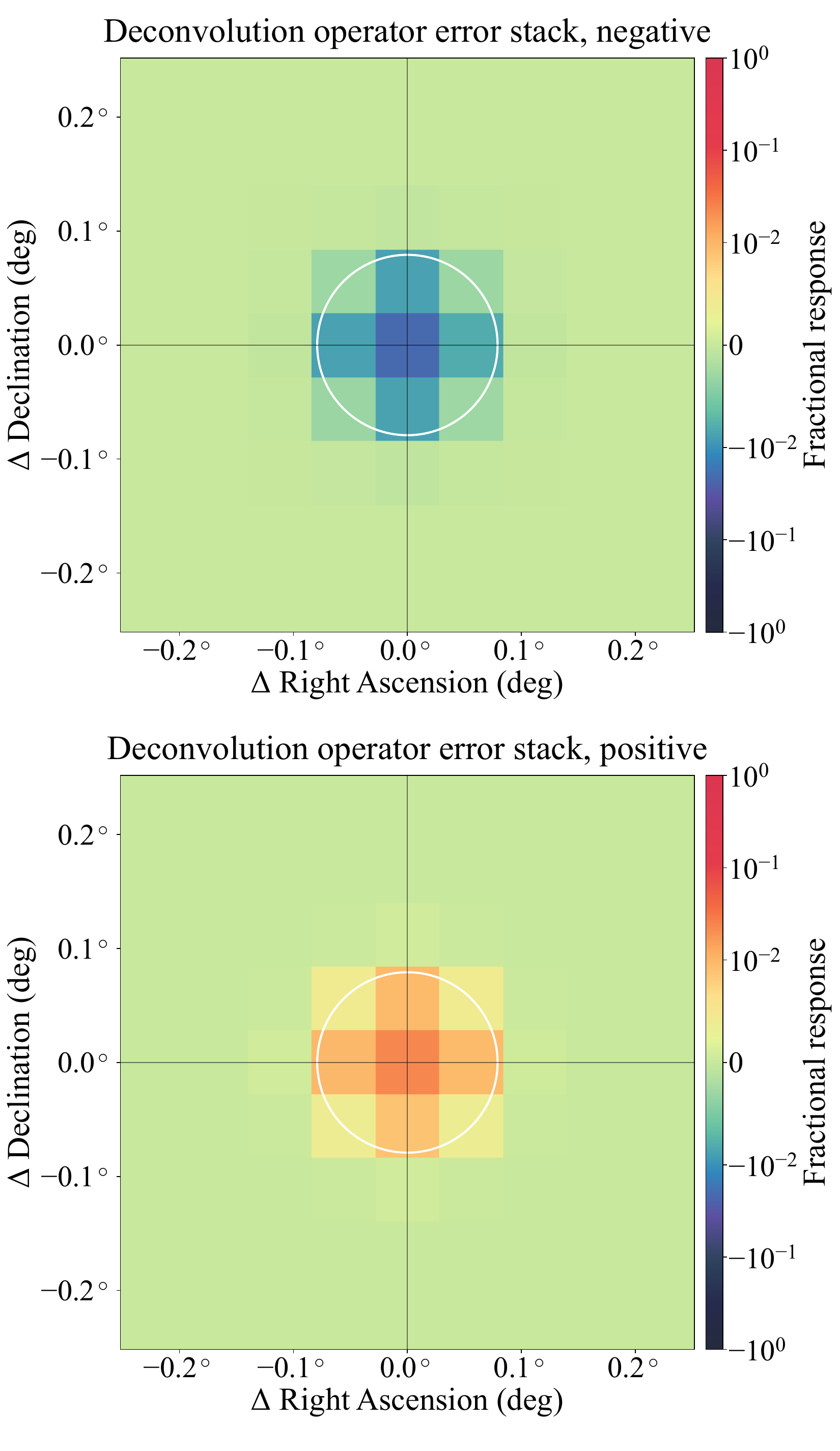}
	\caption{A mean flux density stack of point sources from the bottom panel of Figure~\ref{fig:decon_flux_error} normalized to their input catalog flux, separated by the sign of their corresponding background residual. The top panel corresponds to the point sources that lie in negative residuals, and the bottom panel corresponds to the point sources that lie in positive residuals. There is a 4\% suppression/enhancement of the recovered flux densities, respectively. These values depend on the location and brightness of complex sources which generate the background residuals. The FWHM of the synthesized beam, or resolution, is shown in white.}
	\label{fig:stacked_decon_flux_error}
\end{figure}

We perform a simple stacking experiment using the cross-matched point sources about a primary beam power of 0.5 from the bottom panel of Figure~\ref{fig:decon_flux_error} to illustrate this effect. Sources with positive residual backgrounds and sources with negative residual backgrounds are grouped into separate stacks. The recovered sources are normalized with the flux density cross-matched from the input catalog, and the mean is taken across all normalized sources in the stack. We see in Figure~\ref{fig:stacked_decon_flux_error} that there is approximately 4\% suppression of recovered flux density when the source lies in a negative residual, and approximately 4\% enhanced recovered flux density when the source lies in a positive residual. 

This suppression and enhancement depends solely on the Fornax A residuals. Reducing these residuals by reducing the number of minor cycles, or applications of a global operator $O^\dagger O = O^\dagger_G O_G$, does help, but employing any number of minor cycles, albeit small, will have consequences. For our simulation with Fornax A, using only a few minor cycles per major cycle still results in a 3\% enhanced recovered flux density when a source lies in a positive residual. Negative residual suppression is reduced to minimal levels in this scenario, but this has more to do with the deconvolution algorithm of FHD, which preferentially supports positive components over negative components. 

\section{The Impact of Choosing a Gridding Kernel}
\label{sec:gridding_kernel}

In Section~\ref{sec:now_math}, we outlined the motivation of gridding from a continuous, Hilbert-adjoint perspective. Gridding with the true adjoint, or the $(u,v)$ representation of the Hermitian-conjugate instrumental sensitivity response $\tilde{A}^\dagger(u,v)$, maps the measurements onto a subspace that has the same geometry as the measurement. Mathematically, this means that the span and the metric are preserved; intuitively, this means that the correlations between measurements are preserved. 

However, gridding with the true adjoint may not be practical. Calculating the instrumental sensitivity response is an active area of research \citep{sutinjo_understanding_2015,ninni_electromagnetic_2019,fagnoni_design_2021,sampath_primary_2026}, and can be difficult for certain instruments. Even with knowledge of the instrumental sensitivity response, it may be too complicated (i.e.~\citealt{chokshi_dual_2021,ohara_uncovering_2025,brackenhoff_robust_2025}), and/or difficult to represent in Fourier space \citep{barry_role_2022}. Here, we outline the practicalities of choosing a different gridding operator other than $\tilde{A}^\dagger(u,v)$ using the same operator logic as Section~\ref{sec:now_math}.

As an example, let's split the instrumental sensitivity into the main lobe and the sidelobes,
\begin{equation}
A(l,m) = A_M(l,m) + A_S(l,m).
\end{equation}
Following the logic of Section~\ref{sec:operator_action}, the adjoint operator becomes their linear combination,
\begin{multline}
\label{eq:decomp_operators}
O^\dagger = \tilde{A}^\dagger_M(u-u_n,v-v_n) + \tilde{A}^\dagger_S(u-u_n,v-v_n) \\ = O^\dagger_M + O^\dagger_S.
\end{multline}

Sidelobes are difficult to model and generally faint; some analyses choose to omit them in the gridding adjoint operator altogether. If we make that choice, then the gridding adjoint operator in Equation~\ref{eq:decomp_operators} is just $O^\dagger_M$, even though the act of measuring was still performed through the operator $O_M + O_S$. The normal operator $O^\dagger O$ becomes
\begin{multline}
\label{eq:decomp_normal}
O^\dagger O = O^\dagger_M O_M + O^\dagger_M O_S + O^\dagger_S O_M + O^\dagger_S O_S \\ \xrightarrow{O^\dagger\rightarrow O_M^\dagger} O^\dagger_M O = O^\dagger_M O_M + O^\dagger_M O_S .
\end{multline}
Applying this false adjoint operator during gridding results in the loss of two terms. Their impact on the final space can be assessed through the result of their combined action, 
\begin{equation}
\label{eq:action_loss}
(O^\dagger_S O_M + O^\dagger_S O_S)(\tilde{I'}).
\end{equation}
One important thing to note about the operator loss represented by Equation~\ref{eq:action_loss} is that it is sky-dependent. Whether a gridding operator is appropriate given a desired level of error depends on the sky in question.

\subsection{Simulated example of gridding errors}
\label{subsec:gridding_example}

To give an example of a realistic operator loss, we take the same simulations from Section~\ref{subsec:decon_example} and make ``dirty images" of their gridded visibilities. Again, these simulations only contain a realistic sky $I$ and a realistic primary beam response $A$. The simulations cover two skies: one with no significantly bright sources and one with a complex and bright source right outside the half-power point of the primary beam.

\begin{figure}
\centering
	\includegraphics[width =\columnwidth]{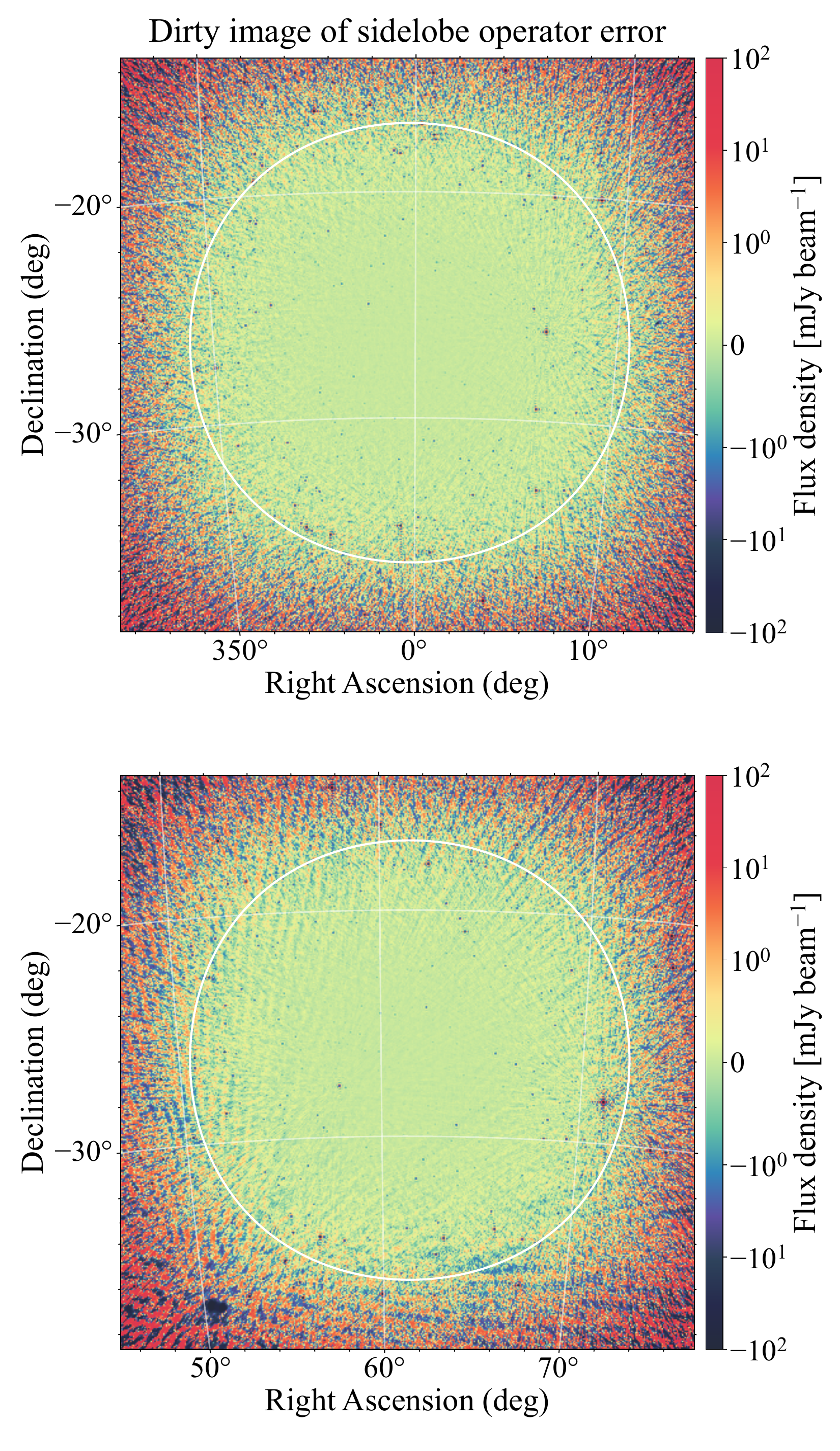}
	\caption{A simulated dirty image of the error when a Gaussian is used as the adjoint gridding operator for the MWA. The top panel is a simulation with no bright sources, and the bottom panel is a simulation with Fornax A at RA$\approx$50$\degree$. Using the simplification of a Gaussian for the adjoint operator artificially concentrates flux density, making the point sources appear brighter by a few mJy/beam. Blue indicates there is more flux density than there should be, and red indicates there is less. The half-power beam level is highlighted in white. }
	\label{fig:gauss_operator_loss_image}
\end{figure}

We use FHD to grid these simulated measurements with the sidelobe gridding adjoint operator, $O^\dagger_S$. This produces the exact operator action loss of Equation~\ref{eq:action_loss} when a Gaussian kernel is used to represent the primary lobe, or $(O^\dagger_S O_M + O^\dagger_S O_S)(\tilde{I'})$. A dirty image of this gridded operator loss is shown in Figure~\ref{fig:gauss_operator_loss_image}. Because no deconvolution or renormalization other than Briggs weighting is applied, Figure~\ref{fig:gauss_operator_loss_image} can be added to a dirty image of the Gaussian-gridded response to create the full reconstruction; it is a linear component. Blue indicates that there is more flux density when a Gaussian adjoint operator is used, and red indicates that there is less flux density. The half-power primary beam level is highlighted in white. 

Figure~\ref{fig:gauss_operator_loss_image} shows that the simplification of the adjoint operator into a Gaussian results in a flux density error around sources. In essence, a Gaussian gridding kernel artificially concentrates flux density in the image plane. This makes point sources appear brighter than they naturally are by a few mJy per beam. By eye, the artificial brightening appears to be more prominent further out in the primary beam, though residuals dominate quickly. 

\begin{figure}
\centering
	\includegraphics[width =\columnwidth]{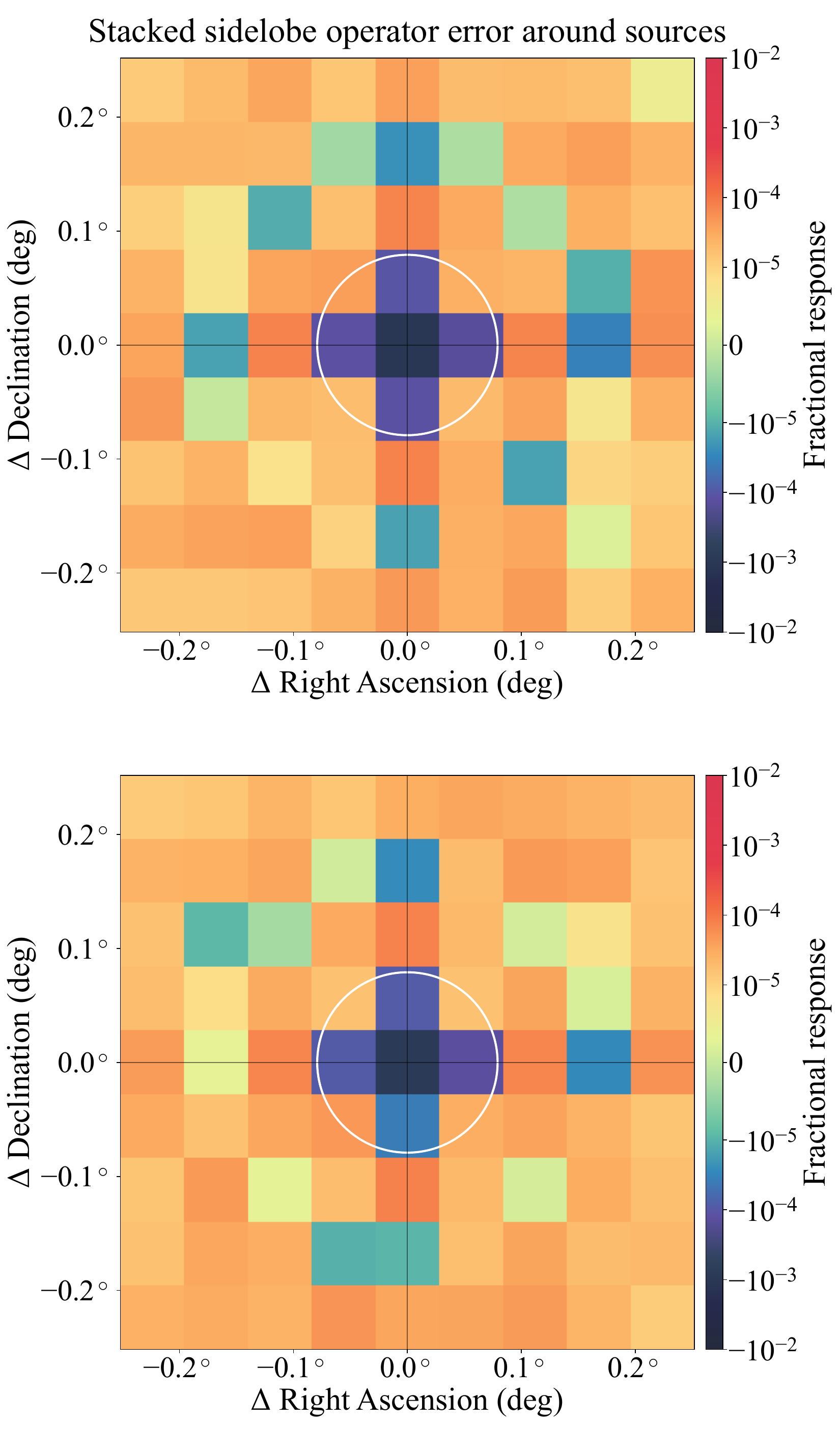}
	\caption{The point-source location stack of the error when a Gaussian is used as the adjoint gridding operator for the MWA, normalized by the true point-source brightness, for each simulation in Figure~\ref{fig:gauss_operator_loss_image}. The error is biased negative, meaning that a Gaussian-gridded image overestimates the per-pixel source brightness by a few parts in 10$^4$. The FHWM of the synthesized beam, or resolution, is shown in white.}
	\label{fig:gauss_operator_loss_stack}
\end{figure}

To illustrate this, Figure~\ref{fig:gauss_operator_loss_stack} is a stack of the image at the location of each point source, normalized by the true flux density of the source, all above the half-power primary beam level. This creates a fractional error caused by the operator loss caused by excluding primary beam sidelobes on the sky. We perform a Bayesian analysis of the central pixel distribution over stacks using a Student-t population model as a prior. Above the half-power primary beam value, over 95\% of point sources exhibit fractional errors exceeding one part in 10$^4$, and about 60\% of point sources exhibit fractional errors exceeding one part in 10$^3$; closer to the center at a primary beam value above 80\% power, these rates drop to 90\% and 10\%, respectively. The inferred response distributions are also heavily skewed negative, as the posterior location parameter $\mu$ is consistently negative regardless of beam power value. 

In other words, gridding with a Gaussian for the MWA \textit{overestimates} the source brightness by a fractional error of roughly a few parts per 10$^4$, dependent on its location within the primary beam. This is a small error, but can become substantial if the source in question is extremely bright.

\section{Functional analysis derivation of gridding \textit{with w}} 
\label{sec:math_with_w}

In all previous sections, we solved for the operator, the adjoint operator, and the operator normal with the assumption that the $w$-term, $W(l,m;w)$,  was negligible. However, it is generally not realistic to have a truly coplanar array with respect to the sky for either large or widefield interferometers. To include $W(l,m;w)$, we can follow the same derivation as in Section~\ref{sec:now_math}, albeit with some important modifications. 

We start our derivation using Equation~\ref{eq:sampled_v_w} -- the observed visibility distribution function with the $w$-component of the exponential separated out into $W(l,m;w)$. Our new representation of the sampling function, $S(u,v,w)$, now includes a delta function in $w$ in addition to those in $u$ and $v$. Thus, our equivalent visibility distribution function (Equation~\ref{eq:scalar_conv_at_point}) now becomes 
\begin{multline}
\label{eq:w_scalar_conv_at_point}
V_{\mathrm{obs}}(u,v,w) = O(\tilde{I}')=\\ \sum_{n=1}^N \left[(\tilde{I'} * \tilde{A} * \tilde{W_n})(u_n,v_n)\right] \delta_{u_n}\delta_{v_n}\delta_{w_n},
\end{multline}
where we have defined $W_n = W(l,m;w_n)$. 

Now, instead of a convolution of two Fourier-transformed functions, we have a convolution of three Fourier-transformed functions. This convolution remains defined in $(u,v)$ space because the measurement is still on a 2D spherical manifold embedded in a 3D space. Our non-coplanar phase screen $W_n$ is defined on the same 2D spherical manifold as the brightness distribution. The inclusion of $\delta(w-w_n)$ therefore only serves to index measurements in $w$ within our distributional space -- essentially, it is selecting a non-coplanar phase screen rather than adding dimensionality. Formally, this means that while we are describing the space as $O : L^2(\mathbb{R}^2) \rightarrow D'(\mathbb{R}^3)$ with the inclusion of $\delta(w-w_n)$, it is also equivalent to describe it as $O : L^2(\mathbb{R}^2) \rightarrow D'(\mathbb{R}^2 \times \mathcal{W})$ where $\mathcal{W}$ is the set of discrete, sampled $w$.

From Equation~\ref{eq:w_scalar_conv_at_point}, we can continue the logic of Section~\ref{sec:now_math} to define the adjoint operator,
\begin{equation}
\label{eq:w_adjoint}
\begin{aligned}
O^\dagger(\delta_{u_m}\delta_{v_m}\delta_{w_m})
&= (\tilde{A}*\tilde{W}_m)^\dagger(u-u_m,v-v_m) \\
&= \tilde{A}^\dagger(u-u_m,v-v_m)*\tilde{W}_m^\dagger(u,v).
\end{aligned}
\end{equation}
We write the operator in two equivalent forms to highlight different interpretations: 1) as the convolution of the Fourier-transformed instrumental sensitivity with the Fourier transform of the $w_m$-dependent kernel, shifted to $(u_m,v_m)$; and 2) as the convolution of the shifted Fourier-transformed instrumental sensitivity with the unshifted Fourier transform of the $w_m$-dependent kernel. Each interpretation results in a different algorithmic approach.

The first interpretation of the adjoint operator in Equation~\ref{eq:w_adjoint} is the fundamental operation in AW-projection (see \citealt{bhatnagar_correcting_2008} for A-projection; see \citealt{cornwell_noncoplanar_2008} for W-projection). AW-projection creates a bespoke gridding operator per instrumental beam sensitivity $\tilde{A}$ and per non-coplanar phase screen $\tilde{W}_n$, usually formed by multiplication in image space. This product is Fourier transformed and Hermitian conjugated to create the phase-screen-corrected adjoint, and then shifted to every corresponding $(u_m,v_m)$ visibility location. The order of the shifting is really what dictates the algorithm as AW-projection. In mathematical terms, this operation is also equivalent to $\widetilde{A\,W_m}^\dagger(u-u_m,v-v_m)$. 

The second interpretation of the adjoint operator in Equation~\ref{eq:w_adjoint} is the core concept behind W-stacking \citep{offringa_wsclean_2014}. For each separate phase screen $\tilde{W}^\dagger_n$, W-stacking builds a gridded $(u,v)$ plane using the shifted adjoint operator $\tilde{A}^\dagger(u-u_m,v-v_m)$, equivalent to the $w=0$ case. Then, that $(u,v)$ plane is Fourier transformed and multiplied with its corresponding Hermitian-conjugated image space phase screen, $W^\dagger_n$. Each separate plane is combined and transformed to create a final, phase-corrected $(u,v)$ plane. In contrast to AW-projection, W-stacking shifts $\tilde{A}^\dagger$ only, keeping $\tilde{W}^\dagger_n$ static. While W-stacking can be done for each individual $\tilde{W}^\dagger_n$, real speed-up comes from discretizing $w$ into layers such that many visibilities can be gridded to the same plane, depending on science goal \citep{offringa_precision_2019}. 

However, the adjoint operator is generally measurement-dependent given the dependence on $w_m$: visibilities with different $w$-values will require different kernels. In mathematical terms, this can be described as one linear operator built from a family of modification kernels indexed by the measurement index $m$, encoded via $\delta_{u_m}\delta_{v_m}\delta_{w_m}$. The normal operator, $O^\dagger O$, is now a convolution kernel of the form
\begin{multline}
\label{eq:generalised_kernel_w}
K(u,v;u',v') = \\ \sum_{n=1}^N (\tilde{A}*\tilde{W}_n)(u_n-u',v_n-v')  (\tilde{A}*\tilde{W}_n)^\dagger(u-u_n,v-v_n),
\end{multline}
where we have arbitrarily chosen to represent the operator as the first representation from Equation~\ref{eq:w_adjoint}. 

The kernel remains four-dimensional like the non-$w$ scenario in Section~\ref{sec:now_math}. Again, this is because we are still acting on a 2D celestial manifold in 3D space. Increasing the dimensionality of the operator (to 3D, as suggested in \citet{cornwell_radio-interferometric_1992}) or kernel (to 6D) would reformulate the geometry of the original measurement equation, and could introduce correlations, power leakage, or false smoothing between $w$ planes that do not exist. We have not gained dimensionality by including $w$, but we have gained a non-stationary operator. In other words, we have not gained depth to objects on the sky, but we have gained a measurement-dependent phase modulation. 

\section{The impact of not including w} 
\label{sec:w_impact}

Omitting the $w$ term during image reconstruction causes decorrelation and distortion in non-coplanar baselines \citep{cornwell_radio-interferometric_1992}. Many algorithms have been built to account for this effect, namely W-projection (i.e. \citealt{cornwell_radio-interferometric_1992,cornwell_noncoplanar_2008}), W-stacking (i.e. \citealt{offringa_wsclean_2014, ye_high_2022}), snapshots (i.e. \citealt{ord_interferometric_2010, offringa_wsclean_2014}, and hybrid methods (i.e. \citealt{pratley_fast_2019,pratley_w-stacking_2020}).

In contrast to Sections~\ref{sec:operator_action} and \ref{sec:gridding_kernel}, we cannot use linearity to describe the operator as a sum of two components, a simplification and a residual, for the whole array. In this case, our full operator contains a convolution with a phase screen that can be unique for each baseline, and thus cannot be linearly decomposed.

Nevertheless, the full form of the operator normal error can be described with 
\begin{multline}
\label{eq:w_operator_loss}
\left(O_W^\dagger O_W-O_A^\dagger O_W\right)(\tilde{I}')
= \\
\left(O_W^\dagger-O_A^\dagger\right)O_W(\tilde{I}'),
\end{multline}
where $O_W$ is the operator with $w$-terms (Equation~\ref{eq:w_adjoint}) and $O_A$ is the operator without $w$-terms (Equation~\ref{eq:adjoint}). As before, the measurement naturally has $w$-terms, or $O_W$, and it's only through our choice of adjoint operator ($O_W^\dagger$ or $O_A^\dagger$) that we make assumptions and thus errors on the image reconstruction. 

\subsection{Simulated example of w-term errors}

To explore the impact of omitting $w$ from the image reconstruction, we choose a conservative example to add to the pre-existing literature. We take our same simulations from Sections~\ref{subsec:decon_example} and \ref{subsec:gridding_example} and reinstate the small physical non-coplanarity of the MWA. Again, these simulations are of two realistic skies, where one sky does not have bright sources and where the other has a bright and complex source just outside the half-power of the primary beam. 

Importantly, these simulations are of two near-zenith transits of 2\,min. The sky rotates overhead during that 2\,min observation. Therefore, the array is not tangent to the sky phase center during the entire observation, but it will be very close. 

We use FHD to do image reconstruction with and without $w$ terms by using the operator adjoint $O_W^\dagger = (\tilde{A}*\tilde{W}_m)^\dagger(u-u_m,v-v_m)$ and $O_A^\dagger = \tilde{A}^\dagger(u-u_m,v-v_m)$, respectively. This algorithm is closest to AW-projection, but in addition propagates gridded weights and variances for use in properly constructed image weights and power spectrum estimates \citep{barry_fhd/eppsilon_2019}. The granularity of the $w$ resolution in binning can be tuned; in the examples provided, the results have reached an asymptote with regards to the $w$ resolution.  

\begin{figure}
\centering
	\includegraphics[width =\columnwidth]{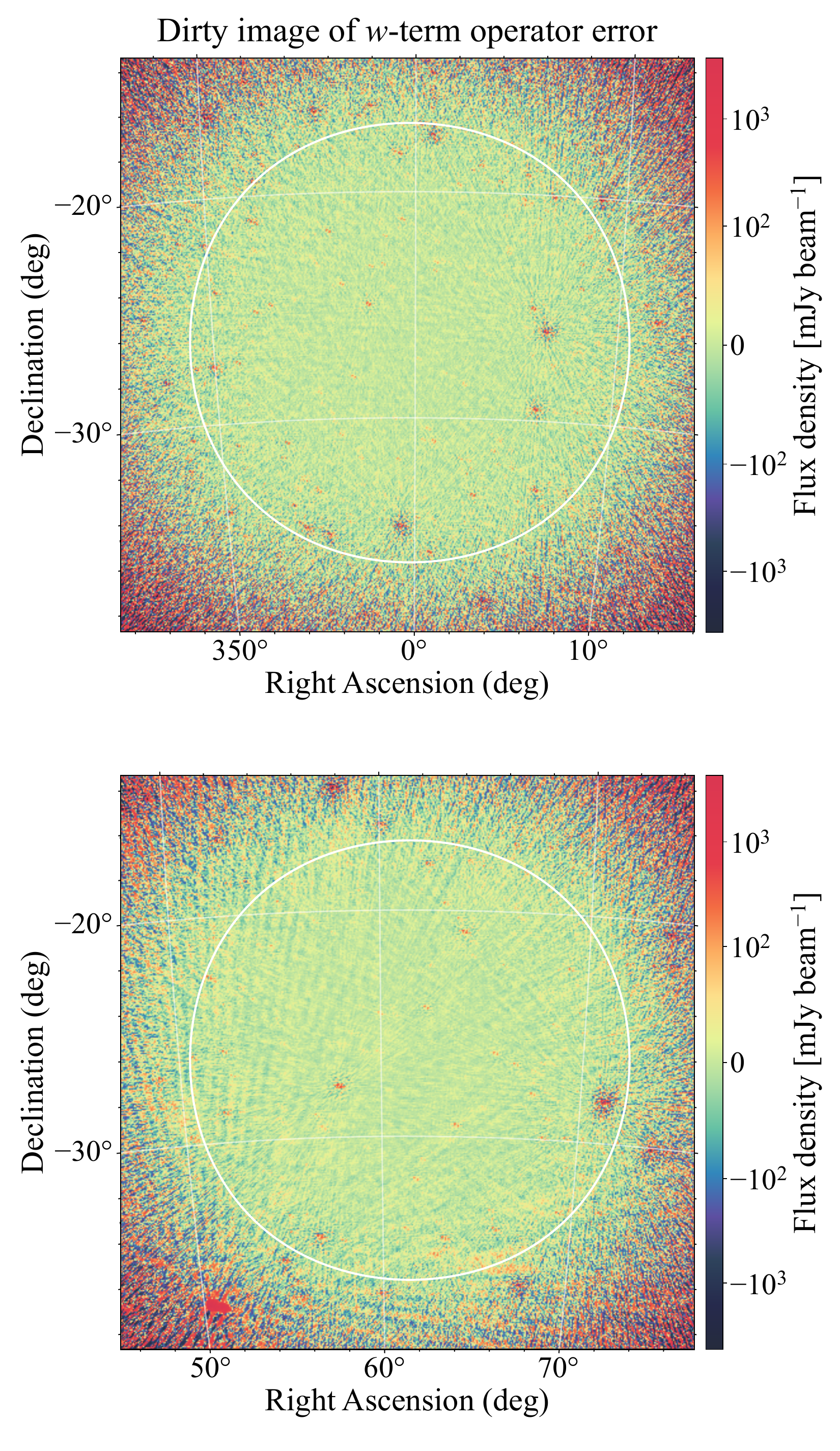}
	\caption{A simulated dirty image of the error when $w$-term, or $W(l,m;w_n)$, is not used as part of the adjoint gridding operator for the MWA. The top panel is a simulation with no bright sources, and the bottom panel is a simulation with Fornax A at RA$\approx$50$\degree$. Using the simplification of $w=0$, or $W(l,m;w_n)=1$ in the adjoint operator artificially decorrelates flux density.  Red indicates there is more flux density when the $w$-term is handled correctly, and blue indicates there is less. The half-power beam level is highlighted in white. }
	\label{fig:w_operator_loss_image}
\end{figure}

We plot the error of not including $w$ terms in Figure~\ref{fig:w_operator_loss_image}. Both fields exhibit the same behavior, with errors concentrated around sources. The half-power of the primary beam is highlighted in white. Upon closer inspection, the error around the sources produces a double-lobed feature. 

\begin{figure}
\centering
	\includegraphics[width =\columnwidth]{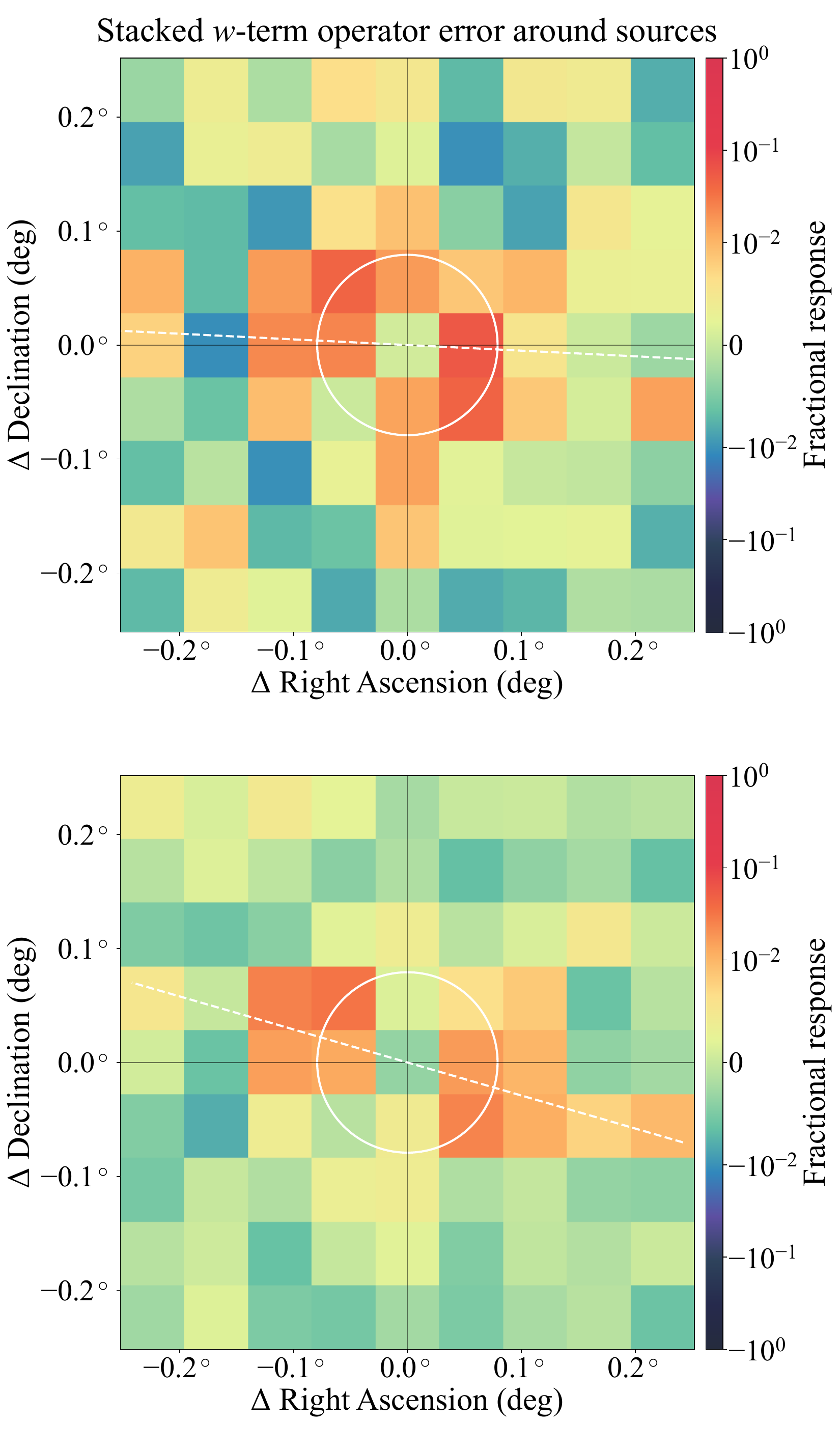}
	\caption{The point-source location stack of the error when the $w$-term is omitted from the adjoint gridding operator for the MWA, normalized by the true point-source brightness, for each simulation in Figure~\ref{fig:w_operator_loss_image}. For a 2\,min observation transiting near-zenith for the mostly co-planar MWA, there is a characteristic line of maximum decoherence dependent on the phase centre location, shown as the white dashed line. When the $w$-term is included, more flux density is recovered along the line of decoherence by upwards of 10\%. The FHWM of the synthesized beam, or resolution, is shown in white.  
    }
	\label{fig:w_operator_loss_stack}
\end{figure}

To highlight this effect, we plot the flux-normalized stack around point sources within the half-power of the primary beam like in Sections~\ref{subsec:decon_example} and \ref{subsec:gridding_example} in Figure~\ref{fig:w_operator_loss_stack}. As before, we include the FWHM of the synthesized beam, or resolution, as a solid white circle. However, we now include a line of expected maximum decoherence, shown as the dashed white line. For each observation, we use a Student-t regression to fit a plane to the $(u,v,w)$ coordinates of the visibilities for each 2\,s time channel, and then average the planes across time. We transform the plane into the image, and plot the maximum gradient in $w$. 

For the bottom panel, it is especially apparent that the observed decoherence, shown as red positive flux density around the source location, follows the line of expected maximum decoherence. The top panel is less coherent due to having a smaller maximum gradient in $w$. Nevertheless, for both stacks, nearly 80\% of the positive response is located within a 45$\degree$ wedge around the line of maximum decoherence. We model the stacks using Gaussian likelihoods and mean-zero Gaussian priors. The resulting posterior distributions for the net fractional difference have posterior means of 10\% and 6\% for the top and bottom stacks, respectively. Along the direction of expected maximum decoherence, the posterior probability of a positive decoherence signal is greater than 0.999.

In other words, sources are up to 10\% less bright in their integrated flux due to $w$-term decoherence for only a 2\,min near-zenith observation for the MWA.  While this has some dependence on the location within the primary beam, this level of error is prevalent throughout the entirety of the image. This difference is due mostly to the change in tangent plane with respect to the phase center rather than the difference in physical altitude changes across the MWA \citep{offringa_wsclean_2014}.

\section{Discussion and Future Directions} 
\label{sec:SH_basis_future}

We outline various ways to extend our functional analysis framework below, both for new research avenues and more complete derivations beyond the scope of this work. 

\subsection{Frequency and time dependence}

Throughout this paper, we have restricted the analysis to a monochromatic and monotemporal case. However, the measurement process described by Equation \ref{eq:RIME} is frequency- and time-dependent. In particular, most radio interferometers are fixed to the ground, where the $(u,v)$ sampling coordinates are in units of wavelength and rotate with the sky. Therefore, the sampling function $S(u,v)$ changes. In addition, the size of the primary beam is proportional to the inverse of frequency. The sky brightness distribution is also a complicated function of frequency, with compact sources having their own spectral indices (see \citealt{condon_cosmological_1984, de_gasperin_radio_2018}), with some sources having more complex peaked structures \citep{callingham_extragalactic_2017}. The Galactic plane and large-scale Galactic synchrotron structure also have a steep spectral index that varies across the sky \citep{de_oliveira-costa_model_2008, zheng_improved_2017}. Importantly, the entire sky brightness distribution is modified by the ionosphere before measurement, creating additional frequency- and time-dependent behavior. 

The practicalities of observation also create potential smearing due to finite time and frequency channelization. All of the above examples of frequency and time dependence will additionally be smeared by channelization, where finer frequency and time resolution will result in less smear. In particular, the primary beam bandwidth smearing could potentially be a large effect, depending on instrumentation (i.e., \citealt{subrahmanyan_radio_2004}). Finally, because the ideal gridding kernel is the Fourier transform of the primary beam response, mutual coupling between dipoles for an aperture array (which leads to direction-dependent amplitude and phase differences) will imply a time-varying gridding kernel when tracking a source. The resulting errors from all of these effects could be described by a functional analysis framework.  

\subsection{Dependency on sampling range, layout, and sky}

The geometric errors that we described in our examples are dependent on both the instrument layout and the extent of the $(u,v)$ plane included, as well as the sky in question. Thus, calculating the expected errors should be driven by forward simulations with a full and realistic instrumental sensitivity in the continuous measurement equation. 

Different science cases will require different ranges, layouts, and sky. For image-based science, the maximum range allowed is generally desired for best resolution. Increasing baseline distance without accounting for the change in the adjoint operator by the $w$ phase screen will result in larger errors, dependent on observation length. This can easily surpass 10\% errors on the true flux density of point sources. The morphology of the decoherence will also depend on the layout and phase center. Image-based science may require more complicated skies than those shown in this work. In this instance, the impact of the normal operator via deconvolution may easily result in errors greater than 5\%, depending on the complexity of sources right outside the half-power primary beam. 

Power spectrum science, conversely, will have restricted $(u,v)$ planes and generally quiet fields, and thus will naively have smaller geometric-based errors than image-based science. However, power spectrum science requires extreme spectral precision. How these errors change as a function of frequency will determine their overall contribution to systematics in the power spectrum space.

\subsection{Mitigation strategies}

By understanding the source of errors in their mathematical context, mitigation strategies can be developed more rigorously. For example, one potential exciting research avenue is to design adjoint operators which make the normal operator easier to invert. Since a simplification in the gridding adjoint operator resulted in small errors of 0.1\% , the adjoint operator can be modified to give a more compact and invertible 4D normal operator without too much geometric error. In any case, pseudo-inverses may be a potential route to invert a normal operator that is not full rank \citep{dillon_fast_2013}.

There are mitigation strategies that exist to use simplified adjoint operators that hold beneficial properties. For example, a 2D Blackman-Harris window in $(u,v)$ space matches to the size-scale of the adjoint operator for the Murchison Widefield Array, but decoheres horizon sources through strong sidelobe suppression \citep{trott_chips:_2016}. This approach is similar to the Tapered Gridded Estimator \citep{choudhuri_visibility-based_2014, choudhuri_tapering_2016} or modified gridding kernel \citep{barry_fhd/eppsilon_2019}, which uses a kernel to suppress signal far outside the half-power of the primary beam. However, this direct signal suppression does not remove the sidelobes of those signals in the main lobe of the primary beam. 

These mitigation strategies need to take into account the science case at hand. For example, coarsely splitting the $w$ phase screen in the adjoint operator may allow for enough reduced error for image-based science, but power spectrum science can suffer greatly \citep{offringa_precision_2019}. Forward simulations are required to assess the impact of a mitigation strategy for a specific science case.

\subsection{Spherical harmonic basis}

In this paper, we have described the operator formalism starting in the Cartesian basis $(l,m)$. However, the natural basis for the celestial brightness distribution is the spherical harmonic basis, described as a Hilbert space $L^2(S^2)$ with respect to the coordinates of colatitude $(\theta)$ and colongitude $(\phi)$. Spherical harmonic functions have been used to study various scalar and vector fields by geophysicists \citep[e.g.,][]{pyshtools}, as well as used widely throughout the Cosmic Microwave Background community, particularly with \texttt{healpix} \citep{healpix,healpy}. 

In the past decade, several spherical harmonic imaging techniques for radio astronomy have emerged, primarily for all-sky imaging with low-frequency widefield aperture arrays.
The $m$-mode transit formalism \citep{Shaw_2014a,Shaw_2014b} has yielded several high-quality all-sky maps \citep{Eastwood_2018,kriele_imaging_2022}. Other imaging approaches include the spherical-wave harmonic transform \citep{Carozzi2015} and the all-sky space-based imaging technique proposed by \citet{Kononov2025}. The former provides a spherical harmonic representation of the measurement in the $(u,v)$ domain, and thus a natural expansion of our functional analysis framework can in principle be applied. 

\section{Conclusions} 
\label{sec:conclusions}

We have built a functional analysis framework using operators for widefield radio interferometer image reconstruction. Starting with the continuous measurement equation in Equation~\ref{eq:RIME}, we derive gridding in analytic form in Equation~\ref{eq:gridding}. The resulting distortion caused by the reconstruction process, or the normal operator, is defined in Equation~\ref{eq:generalised_kernel} as a kernel operator. We then show three examples of how this particular framework can easily describe errors in the image reconstruction process using the linearity of operators. 

We start by investigating the impact of the normal operator through the deconvolution process of image analysis. The dirty sky can be described as the mapping of the true sky via a integral operator in Equation~\ref{eq:dirty_sky}, which allows us to investigate the different assumptions that determine the form of the integral operator. If a flat primary beam is assumed, this integral operator reduces to a global convolution operator of the synthesized beam in Equation~\ref{eq:global_convolution_operator}. This global operator is exactly the foundation behind tradition CLEAN algorithms like H\"ogbom. However, widefield instrumentation cannot be described by the flat-beam approximation, and must use the full four-dimensional normal operator which in non-global and varies across the image. The difference between these two adjoint operators can easily be described using linearity in Equation~\ref{eq:image_diff_decon}.

We input WODEN simulations of the Murchison Widefield Array into Fast Holographic Deconvolution to build the global operator and the non-global operator of Equation~\ref{eq:image_diff_decon} using the Holographic Mapping Function, which is related to the normal operator. We perform two deconvolution tests: 1) Major cycles only, or applying the full normal operator, and 2) major and minor cycles, applying the full normal operator after a few iterations of applying the global convolution operator for efficiency. The act of using minor cycles, no matter how few, will result in image reconstruction errors, especially when bright sources exist outside of the field-of-view. This typically results in errors of $\sim5\%$ of the true flux density of point sources within the half-power beam, as shown in Figure~\ref{fig:decon_flux_error}, and is dependent on the residuals of the bright out-of-field sources, as shown in Figure~\ref{fig:stacked_decon_flux_error}. 

Our second example investigates how the image reconstruction process in general is affected by choice of adjoint operator. The true adjoint operator described in Equation~\ref{eq:adjoint} is the Hermitian-conjugated Fourier-space instrumental sensitivity, shifted to measurement locations in $(u,v)$ space. This is rarely used as the adjoint operator in practice due to the difficulty in describing the instrumental sensitivity. We use the linearity of operators to separate an easier choice, a Gaussian that fits the instrumental sensitivity to half-power, from the rest of the primary beam sidelobe structure in Equation~\ref{eq:decomp_operators}. The resulting error in the image reconstruction can then be described as the left-over terms missed by the Gaussian adjoint in Equation~\ref{eq:action_loss}.

Using the Gaussian decomposition of the primary beam, we image the result of this reconstruction error in Figure~\ref{fig:gauss_operator_loss_image}. The flux density of sources becomes artificially concentrated because measured flux density is reconstructed into a smaller footprint via our choice of approximated adjoint operator. Within the half-power primary beam, this reconstruction error is small, only about $\sim0.1\%$ of the true flux density of point sources as shown in Figure~\ref{fig:gauss_operator_loss_stack}. However, this error is very much dependent on the location within the primary beam, with increasing error further out. 

Our third and final example investigates how the third pseudo-dimension, $w$, affects the image reconstruction process. Within the operator formalism, the observed visibilities are subject to a baseline-dependent phase screen as shown in Equation~\ref{eq:w_scalar_conv_at_point}. The adjoint operator can then be described in Equation~\ref{eq:w_adjoint} as either a Hermitian-conjugated phase-shifted Fourier-space instrumental sensitivity per baseline or a Hermitian-conjugated Fourier-space instrumental sensitivity per baseline, convolved with a Hermitian-conjugated phase shift. This order of operations determines the algorithmic approach of either AW-projection or W-stacking, respectively. Interestingly, the error of omitting $w$ cannot be described well using the linearity of operators, and remains multiplicative as shown in Equation~\ref{eq:w_operator_loss}. 

While an already well-studied error, we add to the existing literature in our third example via an investigation on how much the phase screen contributes to image reconstruction error over the course of a small observation. For the Murchison Widefield Array, there is very little altitude change, and thus the majority of the $w$ phase screen will come from the movement of the phase center with the sky rotation. In just 2\,min in near-zenith observations, there is a suppression of flux density of point sources within the half-power primary beam by $\sim10\%$ due to decorrelation when $w$ is omitted, as shown in Figure~\ref{fig:w_operator_loss_image}. This decorrelation is along the line of greatest change in $w$ during the observation, as shown in Figure~\ref{fig:w_operator_loss_stack}.

Through our three examples, we have illustrated how the functional analysis framework is useful in describing image reconstruction errors in widefield radio interferometers. The thorough derivation provided in this work is meant to provide a foundation for the community to pursue other examples. As discussed in Section~\ref{sec:SH_basis_future}, there are many potential ways to use the framework to describe other image reconstruction errors.  

\begin{acknowledgments}
The authors acknowledge Ruby Byrne, Miguel Morales, and Shintaro Yoshiura for their insightful discussions. The authors acknowledge the Wajarri Yamaji as the traditional custodians of the Murchison Widefield Array site, Inyarrimanha Ilgari Bundara. Dr Barry is the recipient of an Australian Research Council Discovery Early Career Researcher Award (project ID DE240101377) funded by the Australian Government. Dr Barry acknowledges the support of the University of New South Wales Scientia Program. This work was supported by resources awarded under Astronomy Australia Ltd’s merit allocation scheme on the OzSTAR national facility at Swinburne University of Technology. OzSTAR is funded by Swinburne University of Technology and the National Collaborative Research Infrastructure Strategy (NCRIS). AI tools were used to assist with code generation and optimization for the analysis reported in the manuscript. The GAI tool used was ChatGPT. Responsibility for the final manuscript text lies entirely with the authors.
\end{acknowledgments}

\begin{contribution}

Nichole Barry was the main contributor. Other authors assisted in concepts and text.

\end{contribution}

%
\facilities{Murchison Widefield Array}

\software{APLpy \citep{aplpy2012,aplpy2019}, Astropy \citep{astropy:2013, astropy:2018, astropy:2022}, Fast Holographic Deconvolution \citep{sullivan_fast_2012}, Matplotlib \citep{Hunter:2007}, NumPy \citep{harris2020array}, pandas \citep{mckinney2010}, SciPy \citep{2020SciPy-NMeth}}





\bibliography{sample701}{}
\bibliographystyle{aasjournalv7}



\end{document}